\documentclass[twocolumn,showpacs,preprintnumbers,amsmath,amssymb]{revtex4}

\usepackage{graphicx}
\usepackage{dcolumn}
\usepackage{bm}

\begin{document}

\title{Spin-orbit-coupled dipolar Bose-Einstein condensates}
\author{Y. Deng$^{1,2}$, J. Cheng$^{3}$, H. Jing$^{2}$, C.-P. Sun$^{1}$, and S. Yi$^{1}$}
\affiliation{$^1$State Key Laboratory of Theoretical Physics, Institute of Theoretical Physics, Chinese Academy of Sciences, P.O. Box 2735, Beijing 100190, China}
\affiliation{$^{2}$Department of Physics, Henan Normal University, Xinxiang 453007, China}
\affiliation{$^{3}$Department of Physics, South China University of Technology, Guangzhou 510640, China}
\date{\today}

\begin{abstract}
We propose an experimental scheme to create spin-orbit coupling in spin-3 Cr atoms using Raman processes. By employing the linear Zeeman effect and optical Stark shift, two spin states within the ground electronic manifold are selected, which results in a pseudo-spin-$1/2$ model. We further study the ground state structures of a spin-orbit-coupled Cr condensate. We show that, in addition to the stripe structures induced by the spin-orbit coupling, the magnetic dipole-dipole interaction gives rise to the vortex phase, in which a spontaneous spin vortex is formed.

\end{abstract}

\pacs{37.10.Vz, 03.75.Mn}
\maketitle

Over the past few years, there has been rapidly growing interest in engineering Abelian and non-Abelian artificial gauge fields in ultracold atomic gases~\cite{ds1,zhu,liu,spielman,gunter,ds2}. Particularly, the non-Abelian gauge field, or more specifically the spin-orbit (SO) coupling, is of fundamental importance in many branches of physics. Fascinating examples include the quantum spin-Hall effect and the topological insulators in condensed matter physics~\cite{zhang}. With the enormous tunability of the interaction and geometry, ultracold atomic gases may offer a tremendous opportunity for studying exotic quantum phenomena in many-body systems with SO coupling~\cite{stan,stan2,wu1,zhai,ho,sau,xu,wu2,hu}.

In their pioneer experiments, the NIST group have realized the light-induced vector potentials~\cite{lin1}, the synthetic magnetic fields~\cite{lin2}, and the electric forces~\cite{lin3} in ultracold Rb gases through Raman processes~\cite{spielman}, which differs from most dark-state based theoretical proposals~\cite{dalibard} in that the linear Zeeman shift is compensated by the two-photon detuning. More remarkably, they also created a two-component SO-coupled condensate of Rb atoms and observed the phase transition from spatially mixed to separated states~\cite{lin4}. An important ingredient in this experiment is that the quadratic Zeeman shift is employed to separate two desired spin states from the remaining one. Hence, this scheme is inapplicable to atoms without nuclear spin, such as certain isotopes of Cr and Dy, in which the quadratic Zeeman effect is absent.

In this Letter, we propose an experimental scheme to create SO coupling in spin-3 $^{52}$Cr atoms by selecting two internal states from the $J=3$ ground electronic manifold. Similar to the NIST group's scheme, ours also relies on Raman processes. However, we utilize the optical Stark shift to compensate the linear Zeeman shift so that the lowest two levels are near degenerate and well separated from other levels, which leads to a pseudo spin-$1/2$ model. The proposed scheme has the advantages that only a moderate magnetic field strength is required and it also applies to atoms without nuclear spin.

An interesting feature of the Cr atom is that it possesses a large magnetic dipole moment, which makes the scalar Cr condensate an important platform for demonstrating the dipolar effects~\cite{pfau}. Moreover, when an atom's spin degree of freedom becomes available, magnetic dipole-dipole interaction (MDDI) also couples the spin and orbital angular momenta, which is responsible for the Einstein-de Haas effects~\cite{santos,ueda}, the spontaneous demagnetization~\cite{bruno} of the Cr condensate, and the spontaneous spin vortices~\cite{yi,kurn,jzhang} in spinor condensates. Unfortunately, in spin-3 Cr condensates, contact interaction also contains spin-exchange terms which are much larger than the strength of the MDDI~\cite{ho2}. Therefore, the spin vortex phases are yet to be observed. In the pseudo spin-${1}/{2}$ Cr condensate, we show that only the MDDI contains spin-exchange terms and a spontaneous spin vortex is readily observable.

\begin{figure}[tbp]
\centering
\includegraphics[width=2.6in]{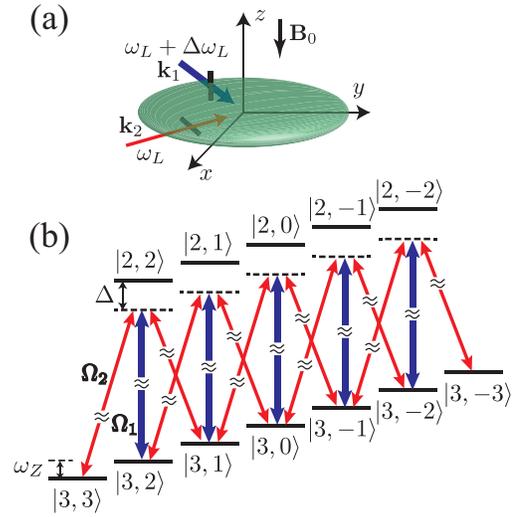}
\caption{(color online). (a) Scheme for creating SO coupling in a Cr atom. Two Raman beams, propagating along $\hat x+\hat y$ and $-\hat x+\hat y$ with frequency difference $\Delta\omega_{L}$, are linearly polarized along $\hat z$ and $\hat x+\hat y$, respectively. A bias field ${\mathbf B}_{0}$ is applied along the negative $z$ axis, which generates a Zeeman shift $\omega_{Z}$ in the ground state manifold. (b) Level diagram for the Raman coupling within the $|J=3\rangle$ ground state manifold by utilizing the $|J'=2\rangle$ excited state.}
\label{scheme}
\end{figure}

We consider a condensate of $^{52}$Cr atoms subjected to a bias magnetic field ${\mathbf B}_{0}$ along the negative $z$-axis. The Zeeman shift within the ground state manifold is $\hbar\omega_{Z}=g_{s}\mu_{B}|{\mathbf B}_{0}|$ with $g_{s}=2$ being the electron spin $g$-factor and $\mu_{B}$ the Bohr magneton. Here, the quadratic Zeeman shift is zero because of the absence of the nuclear spin. As shown in Fig.~\ref{scheme}, atoms are illuminated by a pair of linearly polarized Raman beams which propagate along $\hat x+\hat y$ and $-\hat x+\hat y$ with frequencies $\omega_{L}+\Delta\omega_{L}$ and $\omega_{L}$, respectively. The ground- ($^{7}S_{3}$) to excited-state ($^{7}P_{2}$) transitions are coupled by the Rabi frequencies $\Omega_{1}e^{i{\mathbf k}_1\cdot{\mathbf r}}$ and $\Omega_{2}e^{i{\mathbf k}_2\cdot{\mathbf r}}$, where ${\mathbf k}_{1}=k_{L}(\hat{x}+\hat{y})$ and ${\mathbf k}_{2}=k_{L}(-\hat{x}+\hat{y})$ are the wave vectors of the Raman beams with $k_{L}=\sqrt{2}\pi/\lambda$ and $\lambda$ being the wave length of the lasers. For simplicity, $\Omega_{1,2}$ are assumed to be real. If the frequency of the lasers is far detuned from the ground- to excited-state transition, i.e., $|\Omega_{1,2}/\Delta|\ll 1$  with $\Delta$ being the detuning, the excited states can be adiabatically eliminated to yield the atom-light interaction Hamiltonian

\begin{widetext}
\begin{eqnarray}
\hbar\left( 
{\begin{array}{ccccccc}
U_{2} & \Omega_RX  & U_{2}T &      &   &   &   \\
\Omega_R X^{*} & \Delta_c+U_{1}+U_{2} & \Omega_R(X+X^{*}T) & U_2T & & &  \\
U_2T^{*} & \Omega_R(X^{*}+XT^{*}) & 2\Delta_c+U_{1}+2U_{2} & \Omega_R(X+X^{*}T) & U_2T & & \\
 & U_2T^{*}  & \Omega_R(X^{*}+XT^{*}) & 3\Delta_c+U_{1}+2U_2 & \Omega_R(X+X^{*}T) & U_2T & \\
 & & U_2T^{*} & \Omega_R(X^{*}+XT^{*}) & 4\Delta_c+U_{1}+2U_2 & \Omega_R(X+X^{*}T) & U_2T \\
 & & & U_2T^{*} & \Omega_R(X^{*}+XT^{*}) & 5\Delta_c+U_{1}+U_2 & \Omega_R X^{*}T \\
 & & & & U_2T^{*} & \Omega_RXT^{*} & 6\Delta_c+U_2
\end{array}} 
\right),\label{alint1}
\end{eqnarray}
\end{widetext}
where $\Delta_c=\omega_Z+\Delta\omega_L$ is the two-photon detuning, $\Omega_R=-\Omega_1\Omega_2/\Delta$ is the Rabi frequency for the Raman coupling, $U_{1,2}=-\Omega_{1,2}^2/\Delta$ are the optical Stark shifts induced by the laser fields $\Omega_1$ and $\Omega_2$, respectively, and $T(t)\equiv e^{2i\Delta\omega_L t}$ and $X(x)\equiv e^{2ik_{L}x}$ are introduced for short-hand notation. The physical significance of Eq. (\ref{alint1}) can be readily understood~ \cite{ham1} by using the level diagram [Fig.~\ref{scheme}(b)].

From Hamiltonian (\ref{alint1}), it is apparent that, under the conditions $\Delta_{c}+U_{1}\approx 0$ and $|U_{2}|,|\Omega_{R}|\ll|\Delta_{c}|$, the energy levels $m_J=3$ and $2$ can be separated from other levels due to the large Zeeman shift. These conditions can be satisfied by choosing $\omega_{Z}=\Omega_{1}^{2}/\Delta$ and assuming that $|\Delta\omega_{L}/\omega_{Z}|\ll1$ and $|\Omega_{2}/\Omega_{1}|\ll1$, which eventually leads to an effective two-level Hamiltonian:
\begin{eqnarray}
\hat h =  \frac{{\mathbf p}^2}{2M}\hat I+\hbar\left(
\begin{array}{ccc}
-\Delta\omega_{L}/2 & \Omega_R e^{2ik_{L}x}  \\
\Omega_R e^{-2ik_{L}x} & \Delta\omega_{L}/2
\end{array}
\right), \label{hsingle}
\end{eqnarray}
for pseudo spin-up $|\uparrow\rangle=|m_{J}=3\rangle$ and -down $|\downarrow\rangle=|2\rangle$, where $\hat I$ is the identity matrix and a constant term, $-(U_{2}+\Delta\omega_{L}/2)\hat I$, has been added to obtain Eq. (\ref{hsingle}). We note that the atom-light interaction term in $\hat h$ can be intuitively treated as an effective magnetic field, $${\mathbf B}_{\rm eff}=\hbar(g_{s}\mu_{B})^{-1}(2\Omega_{R}\cos 2k_{L}x,-2\Omega_{R}\sin 2k_{L}x,-\Delta\omega_{L}).$$ Unlike the NIST group's scheme~\cite{lin4}, here, an optical Stark shift $-\Omega_{1}^{2}/\Delta$ is used to compensate the linear Zeeman shift, so that only the levels $m_{J}=3$ and $2$ are Raman coupled near resonance ($\Delta\omega_{L}\approx 0$).

To proceed further, let us focus on the motion of an atom along the $x$ axis by freezing its $y$ and $z$ degrees of freedom. By applying a simple gauge transform~\cite{ho}, the single-particle Hamiltonian can be recast into
\begin{eqnarray}
\hat h_{x}'=\left(\frac{\hbar^{2}q^{2}}{2M}+E_{L}\right)\hat I+2\kappa q\hat\sigma_{z}+\hbar\Omega_{R}\hat\sigma_{x}-\frac{\hbar\Delta\omega_{L}}{2}\hat\sigma_{z},\label{hsinglee}
\end{eqnarray}
where $q=p_{x}/\hbar$ is the quasimomentum, $E_{L}=\hbar^{2}k_{L}^{2}/(2M)$ is the single-photon recoil energy, $\hat\sigma_{x,y,z}$ are the Pauli matrices, and $\kappa=E_{L}/k_{L}$ is the SO coupling strength. Even though $\kappa$ is independent of Raman coupling strength, SO coupling strength is still tunable by varying the relative angle of the Raman beams~\cite{lin4}. It can be readily shown that, after dropping the constant $E_{L}$ term, the eigenenergies of Eq.~(\ref{hsinglee}) are
\begin{eqnarray}
E_{\pm}(q)=\frac{\hbar^{2}q^{2}}{2M}\pm\sqrt{\hbar^{2}\Omega_{R}^{2}+\left(2\kappa q-\frac{\hbar\Delta\omega_{L}}{2}\right)^{2}},\label{disper}
\end{eqnarray}
in analogy to those in the spin-1 Rb condensate. In particular, on the lower branch $E_{-}(q)$, there exist two local minima at $q_{\pm}\simeq\pm k_{L}\sqrt{1-\hbar^{2}\Omega_{R}^{2}/(4E_{L}^{2})}$ when $\hbar\Omega_{R}\lesssim 2E_{L}$ and $\hbar\Delta\omega_{L}\lesssim E_{L}$. The corresponding energies are $E_{-}(q_{\pm})\simeq-E_{L}-\hbar^{2}\Omega_{R}^{2}/(4E_{L})\pm\hbar\Delta\omega_{L}/2$. The states with quasimomenta $\hbar q_{-}$ and $\hbar q_{+}$ (labeled as $|\uparrow'\rangle$ and $|\downarrow'\rangle$, respectively) represent the dressed spin states in which atoms condense in the absence of the interactions.

Here, we would like to discuss the experimental feasibility of our scheme. The transition wavelength from the ground to excited state is $429.1$nm, which corresponds to a recoil energy $E_{L}/\hbar\simeq(2\pi) 10\,{\rm kHz}$. Other laser parameters can be set up as follows. Since the linear Zeeman shift $\omega_{Z}$ in our proposal plays the role of the quadratic Zeeman shift in the NIST experiment~\cite{lin4}, we may set $\hbar\omega_{Z}=3.8E_L$, which implies that the laser intensity $|\Omega_{1}|^{2}=3.8E_{L}|\Delta|/\hbar$ is about the same order of magnitude as that used in the experiment. To allow the Raman coupling $\Omega_{R}$ to vary from $0$ to $E_{L}$, which covers the most interesting parameter region in the experiment, the maximum value of $|\Omega_{2}|$ can be chosen as $0.26|\Omega_{1}|$. Consequently, the maximum value of $U_{2}$ is less than $0.26E_L$, which justifies the neglecting of $U_{2}$ in Eq. (\ref{alint1}). Finally, we point out that the SO coupling strength $\kappa$ in our scheme is $3.14$ times larger than that in the Rb experiment due to the smaller mass and the shorter transition wavelength of Cr atom.

Now we turn to study the many-body effect in a SO-coupled Cr condensate. To this end, we first write down the single-particle Hamiltonian, which, in the second quantized, takes the form
\begin{eqnarray}
\hat H_{0}=\int d{\mathbf r}\hat\Psi^{\dag}({\mathbf r})\left[\hat h+V({\mathbf r})-\mu\right]\hat\Psi({\mathbf r}),
\end{eqnarray}
where $V({\mathbf r})=M\omega_{\perp}^{2}(x^{2}+y^{2}+\gamma^{2}z^{2})/2$ is an axially symmetric harmonic trap with $\omega_{\perp}$ being the radial trap frequency and $\gamma$ the trap aspect ratio, $\mu$ is the chemical potential, and $\hat\Psi({\mathbf r})=[\hat\psi_{\uparrow}({\mathbf r}),\hat\psi_{\downarrow}({\mathbf r})]^{T}$ is the field operator for the bare spin states. We note that $\hat H_{0}$ can also be expressed in terms of dressed spin states by using the transform $\hat\psi_{\uparrow}({\mathbf r})\simeq\hat\psi_{\uparrow'}({\mathbf r})-\varepsilon e^{2ik_{L}x}\hat\psi_{\downarrow'}({\mathbf r})$ and $\hat\psi_{\downarrow}({\mathbf r})\simeq-\hat\psi_{\downarrow'}({\mathbf r})+\varepsilon e^{-2ik_{L}x}\hat\psi_{\uparrow'}({\mathbf r})$, where $\varepsilon\simeq\hbar\Omega_{R}/(4E_{L}+\hbar\Delta\omega_{L})\ll 1$ in the weak Raman coupling limit $\hbar\Omega_{R}/E_{L}\ll1$.

In terms of the bare spin states, the collisional interaction takes the form
\begin{eqnarray}
\hat H_{c}&=&\frac{1}{2}\int d{\mathbf r}\left(g_{6}\hat \psi_{\uparrow}^{\dag}\hat \psi_{\uparrow}^{\dag}\hat \psi_{\uparrow}\hat \psi_{\uparrow}+\frac{5g_{4}+6g_{6}}{11}\hat \psi_{\downarrow}^{\dag}\hat \psi_{\downarrow}^{\dag}\hat \psi_{\downarrow}\hat \psi_{\downarrow}\right.\nonumber\\
&&\left.\qquad\qquad+2g_{6}\hat \psi_{\uparrow}^{\dag}\hat \psi_{\downarrow}^{\dag}\hat \psi_{\downarrow}\hat \psi_{\uparrow}\right),\label{hcol}
\end{eqnarray}
where $g_{4,6}=4\pi\hbar^{2}a_{4,6}/M$ with $a_{4}=58\,a_{B}$ and $a_{6}=112\,a_{B}$ being the $s$-wave scattering lengths for the collisional channel with total spin angular momentum $j=4$ and $6$, respectively~\cite{pfau2}. This result can be understood as follows. The collision between two $m_{J}=3$ atoms can happen only in the total spin $j=6$ channel; consequently, it has a scattering length $a_{6}$. For collisions between $m_{J}=3$ and $2$ atoms, the projection of the total spin along the $z$-axis is $m_{j}=5$, which is conserved during collision. Therefore, this collision also happens in the $j=6$ channel. But when two $m_{J}=2$ atoms collide with each other, both $j=6$ and $4$ channels will contribute. Apparently, the spin-up and -down states are immiscible.

The MDDI for the pseudo spin-$1/2$ system can be decomposed into $\hat H_{d}=\hat H_{d}^{(1)}+\hat H_{d}^{(2)}$ with
\begin{eqnarray}
&&\hat H_{d}^{(1)}=g_{d}\sqrt{\frac{4\pi}{5}}\!\int\!\frac{d{\mathbf r}d{\mathbf r}'}{|{\mathbf r}-{\mathbf r}'|^{3}}
Y_{2,0}(\hat e)\left[-9\hat \psi_{\uparrow}^{\dag}\hat \psi_{\uparrow}'^{\dag}\hat \psi_{\uparrow}'\hat \psi_{\uparrow}\right.\nonumber\\
&&\left.\;\; -4\hat \psi_{\downarrow}^{\dag}\hat \psi_{\downarrow}'^{\dag}\hat \psi_{\downarrow}'\hat \psi_{\downarrow}-12\hat \psi_{\uparrow}^{\dag}\hat \psi_{\downarrow}'^{\dag}\hat \psi_{\downarrow}'\hat \psi_{\uparrow}+3\hat \psi_{\uparrow}^{\dag}\hat \psi_{\downarrow}'^{\dag}\hat \psi_{\uparrow}'\hat \psi_{\downarrow}\right],\\
&&\hat H_{d}^{(2)}=-g_{d}\sqrt{\frac{9\pi}{5}}\!\int\!\frac{d{\mathbf r}d{\mathbf r}'}{|{\mathbf r}-{\mathbf r}'|^{3}}\left[2Y_{2,-1}(\hat e)\left(3\hat \psi_{\uparrow}^{\dag}\hat \psi_{\uparrow}'^{\dag}\hat \psi_{\uparrow}'\hat \psi_{\downarrow}\right.\right.\nonumber\\
&&\left.\left.\;\;+2\hat \psi_{\uparrow}^{\dag}\hat \psi_{\downarrow}'^{\dag}\hat \psi_{\downarrow}'\hat \psi_{\downarrow}\right)+\sqrt{6}Y_{2,-2}(\hat e)\hat \psi_{\uparrow}^{\dag}\hat \psi_{\uparrow}'^{\dag}\hat \psi_{\downarrow}'\hat \psi_{\downarrow}+h.c.\right],\label{hdip}
\end{eqnarray}
here, $g_{d}=\mu_{0}g_{s}^{2}\mu_{B}^{2}/(4\pi)$ with $\mu_{0}$ being the vacuum permeability and $\mu_{B}$ the Bohr magneton, $\hat e=({\mathbf r}-{\mathbf r}')/|{\mathbf r}-{\mathbf r}'|$ is an unit vector, and we have adopted the notations $\hat \psi_{\alpha}\equiv\hat \psi_{\alpha}({\mathbf r})$ and $\hat \psi_{\alpha}'\equiv\hat \psi_{\alpha}({\mathbf r}')$ with $\alpha=\uparrow$ and $\downarrow$. The first three terms of $\hat H_{d}^{(1)}$ represent the intra- and interspecies dipolar interactions in a mixture of $m_{J}=3$ and $2$ atoms, and the last term is the exchange dipolar interaction. $\hat H_{d}^{(2)}$ is of particular interest. It represents the SO coupling containing in the MDDI and does not conserve the atom number in the individual spin state. However, the total angular momentum is conserved by $\hat H_{d}^{(2)}$. 

In this spin-$1/2$ model, interactions have a much simpler form compared to those in the spin-3 system. In particular, here, only the MDDI contains spin-exchange terms. As will be shown, even though $g_{d}$ is much smaller than $g_{4,6}$, the spin associated dipolar effect can be readily detected in pseudo spin-${1}/{2}$ Cr condensates.

\begin{figure}[tbp]
\centering
\includegraphics[width=3.3in]{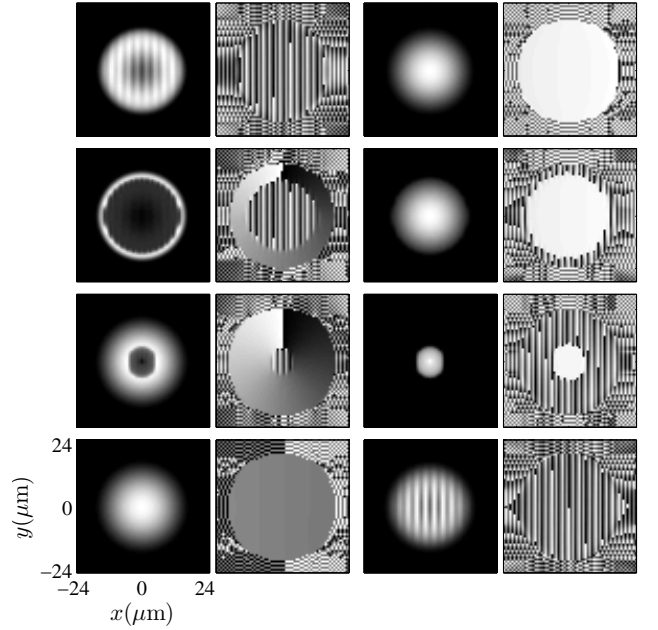}
\caption{Integrated densities (columns 1 and 3 for spin-up and -down, respectively) and phases of the condensate wave functions on the $z=0$ plane (columns 2 and 4 for spin-up and -down, respectively). From the first to the fourth rows, the frequency differences are $\hbar\Delta\omega_{L}/E_{L}=0.01$, $0.04$, $0.0875$, and $0.1$, respectively.}
\label{denpha}
\end{figure}

We now investigate the ground state structures of the SO-coupled dipolar condensate using the mean-field theory. To this end, the field operators $\hat\psi_{\alpha}$ are replaced by the condensate wave function $\psi_{\alpha}=\langle\hat \psi_{\alpha}\rangle$, which can be obtained by numerically minimizing the free energy functional ${\cal F}[\psi_{\uparrow},\psi_{\downarrow}]=\langle \hat H_{0}+\hat H_{s}+\hat H_{d}\rangle$. Specifically, we consider a Cr condensate with $N=10^{6}$ atoms. The parameters for the trapping potential are chosen as $\omega_{\perp}=(2\pi)100\,{\rm Hz}$ and $\gamma=6$, representing a three-dimensional pancake-shaped trap. Furthermore, the Rabi frequency for Raman coupling is fixed at $\hbar\Omega_{R}=-0.01E_{L}$. Since $\varepsilon\ll1$ is satisfied, we shall discuss only the ground state in terms of the bare spin states.

\begin{figure}[tbp]
\centering
\includegraphics[width=1.8in]{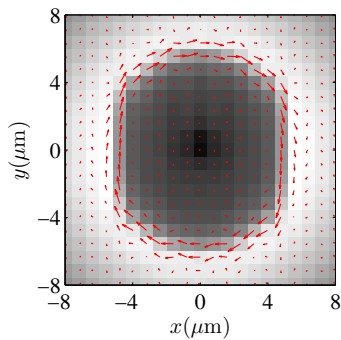}
\caption{(color online). Vector plot of the transverse components of ${\mathbf s}({\mathbf r})$ on the $z=0$ plane for $\hbar\Delta\omega_{L}=0.0875E_{L}$. The grayscale indicates the integrated density $\bar n_{\uparrow}(x,y)$.}
\label{magspn}
\end{figure}

In Fig. \ref{denpha}, we plot the integrated density, $\bar n_{\alpha}(x,y)=N^{{-1}}\int dz|\psi_{\alpha}({\mathbf r})|^{2}$, and the phases of the condensate wave functions for various $\Delta\omega_{L}$'s. When $\Delta\omega_{L}$ is small (the first row in Fig.~\ref{denpha}), the single-particle energies of the two pseudo spin states are nearly degenerate such that the ground state structure is mainly determined by the interactions. Apparently, both $\hat H_{c}$ and $\hat H_{d}$ (in the pancake-shaped trap) favor the spin-down state. Therefore, $|\downarrow~\rangle$ becomes dominantly populated, which we refer to as the {\em polarized phase} (PP). As shown in the fourth row of Fig.~\ref{denpha}, the PP also occurs when the frequency difference $\Delta\omega_L$ (or, equivalently, the $z$ component of the effective magnetic field ${\mathbf B}_{\rm eff}$) is sufficient large, under which $|\uparrow\rangle$ becomes dominantly occupied. 

\begin{figure}[tbp]
\centering
\includegraphics[width=3.3in]{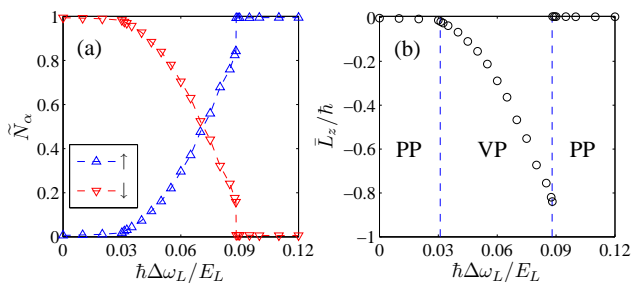}
\caption{(color online). Reduced atom number $\widetilde N_{\alpha}$ (a) and average orbital angular momentum $\bar L_{z}$ (b) as functions of $\Delta\omega_{L}$.}
\label{result}
\end{figure}

The common feature of the PPs is that the wave function of the highly populated state is structureless, as any structure developed in the high density spin state would cost too much kinetic energy. On the other hand, a striped structure forms in both the density and phase of the less populated spin state. The phase stripe can be intuitively understood as follows. To lower the energy, the pseudo spin density, ${\mathbf s}({\mathbf r})=\sum_{\alpha\beta}\psi_{\alpha}^{*}\hat{\boldsymbol\sigma}_{\alpha\beta}\psi_{\beta}$, has to be antiparallel to the local effectively magnetic field, which requires the relative phase of the condensate wave function to take the form ${\rm arg}(\psi_{\uparrow})-{\rm arg}(\psi_{\downarrow})\sim \pi+2k_{L}x$. Since the phase of the highly populated state is a constant, the phase of the other spin state is then periodically modulated along the $x$ direction. The density stripe in the less populated spin state is caused by the immiscible nature of the two-component condensate.

More remarkably, we observe a {\em vortex phase} (VP) for intermediate $\Delta\omega_{L}$ values. As shown in the second and third rows of Fig.~\ref{denpha}, a singly quantized vortex appears in the spin-up state due to the SO coupling in the $\hat H_{d}^{(2)}$ term of the MDDI. In the VP, the atom numbers in spin-up and -down states become comparable [Fig.~\ref{result}(a)], which allows the spin ${\mathbf s}({\mathbf r})$ of the atom to form significant transverse components. As shown in Fig.~\ref{magspn}, since the MDDI is minimized with a head-to-tail spin configuration, the transverse components of ${\mathbf s}({\mathbf r})$ are forced to form a spin vortex. Consequently, the wave function $\psi_{\uparrow}$ develops a $2\pi$ phase winding, representing a vortex state. The reason that the vortex state appears only on the spin-up component is due to the immiscibility of our two-component system, which results a density depletion at the center of $\psi_{\uparrow}$. Therefore, forming a vortex in the spin-up state costs less kinetic energy. Moreover, in the VP, the phase stripes also appear in the low density regions of both spin states, which is the manifestation of the SO coupling induced by the light fields.

To determine the phase boundaries, we plot the $\Delta\omega_{L}$ dependences of the reduced atom number, $\widetilde N_{\alpha}=N^{-1}\int d{\mathbf r}|\psi_{\alpha}|^{2}$, and the average orbital angular momentum, $\bar L_{z}=N^{-1}\sum_{\alpha}\int d{\mathbf r}\psi_{\alpha}^{*}\hat L_{z}\psi_{\alpha}$, in Fig.~\ref{result}, where $\hat L_{z}=-i\hbar(x\frac{\partial}{\partial y}-y\frac{\partial}{\partial x})$ is the $z$-component of the orbital angular momentum. As can be seen, the phase boundaries  are marked by two critical $\Delta\omega_{L}$ values, $\Delta\omega_{L}^{*}=0.031E_{L}/\hbar$ and $\Delta\omega_{L}^{**}=0.088E_{L}/\hbar$. For $\Delta\omega_{L}^{*}<\Delta\omega_{L}<\Delta\omega_{L}^{**}$, the condensate lies in the VP; otherwise, it is in the PP. Within the VP, atom numbers and orbital angular momentum change dramatically as one varies $\Delta\omega_{L}$.

In conclusion, we have proposed an experimental scheme to generate SO coupling in spin-3 Cr condensates via Raman processes. Optical Stark shift is employed to selecting two spin states from the atom's ground electronic manifold. The proposed scheme should be readily realizable experimentally. Subsequently, the ground-state structures of a SO-coupled Cr condensate have been investigated. We show that the interplay between the light fields and the MDDI gives rise to the polarized and vortex phases. In particular, a spontaneous spin vortex is formed in VP. The spin vortex state can be experimentally identified if it is observed that the spin-up condensate is a vortex state and the spin-down condensate is a vortex-free one. Finally, we point out that our scheme should also apply to the Dy atom~\cite{dybec}, which has an even larger dipole moment.

We thank Ruquan Wang for helpful discussions. This work was supported by the NSFC (Grants No. 11025421, No. 11174084, and No. 10935010) and National 973 program (Grants No. 2012CB922104 and No. 2012CB921904).


\begin{thebibliography}{99}
\bibitem{ham1}  For example, considering the diagonal element corresponding to $|3,1\rangle$, as shown in Fig.~\ref{scheme}(b), it couples to $|2,2\rangle$, $|2,1\rangle$, and $|2,0\rangle$ via the laser fields $\Omega_2$, $\Omega_1$, and $\Omega_2$, respectively. Consequently, we have the light shifts $U_1+2U_2$ besides the laser detuning $2\Delta_c$. As another example, the coupling between $|3,1\rangle$ and $|3,0\rangle$ is caused by two Raman processes: $|3,1\rangle \rightarrow |2,0\rangle \rightarrow |3,0\rangle$ and $|3,1\rangle \rightarrow |2,1\rangle \rightarrow |3,0\rangle$, which generate the terms $\Omega_{R}X$ and $\Omega_{R}X^{*}T$, respectively. The time-dependence, $T$, in the latter process is due to the energy mismatch, while the space-dependencies, $X$ and $X^{*}$, are caused by the momentum mismatch. Finally, the $U_{2}T$ terms correspond to the Raman process $|3,m_{J}\rangle\rightarrow|2,m_{J}-1\rangle\rightarrow|3,m_{J}-2\rangle$.
\end{thebibliography}
\end{document}